\documentclass[12pt,a4paper]{article}
\usepackage{a4wide}
\usepackage{graphicx}
\usepackage{epsfig}
\usepackage{cite}
\usepackage{amssymb}
\usepackage{amsfonts}
\usepackage{amsmath}

\def    \beq           	{\begin{equation}}
\def    \eeq           	{\end{equation}}
\def    \bea           	{\begin{eqnarray}}
\def    \eea           	{\end{eqnarray}}
\def    \nn            	{\nonumber}  
\def    \lraw           {\leftrightarrow}
\def    \raw           	{\rightarrow}

\def	\cN		{{\mathcal{N}}}

\def	\cO		{{\mathcal{O}}}
\def	\cW		{{\mathcal{W}}}
\def 	\bZ		{{\mathbb Z}}

\def	\bR		{{\mathbb R}}

\newcommand{\id}{1\!\!1}

\renewcommand{\thefootnote}{\fnsymbol{footnote}}

\begin{document}

\title{
\begin{flushright}
\ \\*[-80pt]
\begin{minipage}{0.2\linewidth}
\normalsize
KUNS-2086\\*[50pt]
\end{minipage}
\end{flushright}
{\Large \bf 
Three-Family Models from a Heterotic Orbifold\\
on the $E_6$ Root Lattice
\\*[20pt]}}

\author{
Kei-Jiro Takahashi\footnote{
E-mail address: keijiro@gauge.scphys.kyoto-u.ac.jp}\\*[20pt]
{\it \normalsize
Department of Physics, Kyoto University,
Kyoto 606-8502, Japan} \\
 \\*[50pt]}

\date{
\centerline{\small \bf Abstract}
\begin{minipage}{0.9\linewidth}
\medskip
\medskip
\small
We classify $\cN=1$ orbifolds on the $E_6$ root lattice  
and investigate explicit model constructions 
on the ${\mathbb Z}_3 \times {\mathbb Z}_3$ orbifold in heterotic string theory. 
Interestingly some of the twisted sectors from 
the ${\mathbb Z}_3 \times {\mathbb Z}_3$ orbifold 
on the $E_6$ root lattice have just three fixed tori respectively, 
and generate three degenerate massless states. 
We also found three point functions with flavor mixing terms. 
We assume only non-standard gauge embeddings and 
find that they lead to three-family $SU(5)$ and $SO(10)$ GUT-like models. 
These models also include strongly coupled sectors in the low energy and 
messenger states charged with both hidden and visible sectors. 
\end{minipage}
}

\begin{titlepage}

\maketitle

\thispagestyle{empty}
\end{titlepage}

\renewcommand{\thefootnote}{\arabic{footnote}}
\setcounter{footnote}{0}

\section{Introduction}
Superstring theory is one of the most promising candidates for a unified theory. 
To establish a connection between superstring theory in $d=10$ 
and an effective theory in $d=4$, 
it is important to understand the geometries of six-dimensional compact spaces.
Compactification on Calabi-Yau three-folds gives $\cN=1$ models 
\cite{Greene:1986ar,Bouchard:2005ag,Braun:2005nv,Blumenhagen:2006ux},  
but they allow explicit calculations only in a limited number of cases, 
and realistic model building is still difficult. 
An orbifold can be interpreted as a special case of a Calabi-Yau manifold, 
and orbifold compactifications of heterotic string theory 
\cite{Dixon:1985jw,Dixon:1986jc} 
are simple enough to allow investigation of phenomenological properties, 
such as the Yukawa coupling and selection rules of the superpotential. 
Many three-family supersymmetric models, which are 
based on the abelian discrete groups ${\mathbb Z}_N$ and 
${\mathbb Z}_N \times {\mathbb Z}_M$, have been constructed 
\cite{Ibanez:1987sn,Bailin:1987xm,Font:1988mm,Casas:1988hb,Ibanez:1987pj,Ibanez:1987xa,Font:1989aj,Font:1988mk,Forste:2004ie,Forste:2005rs,Forste:2005gc,Kobayashi:2004ud,Kobayashi:2004ya,Buchmuller:2004hv,Buchmuller:2005jr,Buchmuller:2006ik,Nilles:2006np,Giedt:2001zw,Kim:2006hv,Kim:2006hw}. 
The Yukawa couplings can be determined with world sheet instanton techniques, 
and they depend on the geometry of the target space. 
Due to restrictions from the coupling selection rules, 
it is often difficult to obtain realistic mass matrices with flavor mixing terms. 
These orbifold models are classified by use of Coxeter elements, 
which are subgroups of the Weyl group of Lie algebra \cite{Schellekens:1987ij,Kobayashi:1990fx,Bailin:1999nk}. 
For the ${\mathbb Z}_N \times {\mathbb Z}_M$ orbifold models 
constructed to this time, 
the compact spaces are factorizable to $T^2 \times T^2 \times T^2$. 
Recently, an $E_8\times E_8^\prime$ heterotic orbifold on a non-factorizable torus  
has been investigated \cite{Donagi:2004ht,Faraggi:2006bs,Forste:2006wq,Takahashi:2007qc}\footnote{Application to 
non-factorizable orbifolds for TypeIIA models is also possible 
(see Refs. \cite{Blumenhagen:2004di} and \cite{Forste:2007zb}).}.
It is observed that the selection rules for three-point functions are changed 
non-trivially in these orbifolds \cite{Forste:2006wq}, 
and numbers of generations are reduced. 
Thus, there are possibilities to construct more realistic models 
with flavor mixing terms and fewer extra matter fields. 
In Ref. \cite{Takahashi:2007qc}, we classified orbifolds on non-factorizable tori 
according to the Weyl reflections and outer automorphisms of Lie root lattices, 
except orbifolds on the $E_6$ root lattice. 
In this paper, we give all the allowed orbifolds on the $E_6$ root lattice, 
which lead to $\cN=1$ supersymmetric models, 
and complete the classification of the orbifolds on the Lie root lattices. 
A non-factorizable torus on the $E_6$ lattice is highly symmetric 
from the six-dimensional viewpoint. 
A torus on the $E_6$ lattice allows 
the ${\mathbb Z}_3 \times {\mathbb Z}_3$ orbifold, and, interestingly, 
it contains twisted sectors localized at the three fixed tori.
Fot this reason, 
models based on it can naturally lead to three generations of matter. 
In addition, 
we investigate the general structure of the interactions of the orbifold, 
and find that some interactions allow flavor mixing terms, 
where world sheet instantons 
on the ${\mathbb Z}_3 \times {\mathbb Z}_3$ orbifolds 
generate three-point functions with exponentially 
suppressed factors \cite{Hamidi:1986vh}. 

In Section 2 we explain the details of an orbifold on the $E_6$ root lattice.
Section 3 gives examples of three-family GUT-like models and 
general consideration of the three-point interactions.

\section{Orbifolds on an $E_6$ root lattice}
We compactify six dimensions to a torus $T^6$ 
in order to construct four dimensional models.
$T^6$ is obtained by compactifying ${\mathbb R}^6$ 
on a lattice $\Lambda$: 
\beq
T^6 = {\mathbb R}^6/\Lambda.
\eeq
Points in ${\mathbb R}^6$ differing by a lattice vector $L \in \Lambda$ 
are identified as $x \sim x + L$. 
An orbifold is defined to be the quotient of a torus 
over a discrete set of the isometries of the torus, 
called the point group $P$, i.e.
\begin{equation}
\cO = T^6/P={\mathbb R}^6/S. 
\end{equation}
Here, $S$ is called the space group, and it is the semidirect product of the 
point group $P$ and the translation group. 
We consider the cases which $\Lambda$ is generated 
from the $E_6$ root lattice.\footnote{Heterotic orbifolds on Lie root lattices 
($A_n, B_n, C_n, D_n, F_4$ and $G_2$) are classified in Ref. \cite{Takahashi:2007qc}. 
The orbifolds on Lie root lattices would be 
generic toroidal orbifold in six dimensions,
because the other lattices could be realized by continuous deformation of geometric moduli of orbifolds on the Lie root lattices. 
There are interesting coincidence between the non-factorizable models and 
the factorizable models with generalized discrete torsion \cite{Ploger:2007iq}.} 

For a Lie group of rank $l$, 
the Lie root lattice is given by 
\begin{equation}
\Lambda \equiv \left\{ \sum_{i=1}^l n_i \alpha_i | n_i \in \bZ \right\}, 
\end{equation}
where $\alpha_i$ is a simple root. 
The point group of an orbifold must be automorphic of the lattice.
The automorphisms of the Lie root lattice can be generated by 
the Weyl reflections and the outer automorphisms.
The Weyl group $\cW$ is generated by the Weyl reflections: 
\begin{equation}
r_{k}:\lambda \raw \lambda -2\frac{\alpha_k \cdot \lambda}{\alpha_k \cdot \alpha_k} \alpha_k. 
\end{equation}
\begin{figure}[tbp]
\begin{center}
\includegraphics{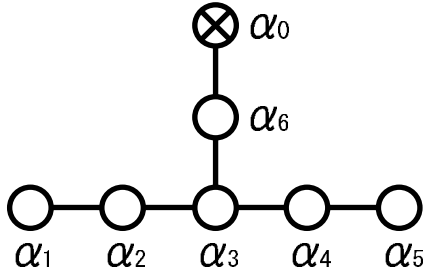}
\caption{\it Extended $E_6$ diagram}
\label{e6}
\end{center}
\end{figure}
The point group, $\theta$ and $\phi$, 
of a ${\mathbb Z}_N \times {\mathbb Z}_M$ orbifold 
on a Lie root lattice can be defined by 
two commutative elements from the Weyl group and the outer automorphisms 
$G_{out}$, i.e. $\theta,~\phi \in \{ \cW,G_{out} \}$.
We consider the case in which $\Lambda$ is
the $E_6$ root lattice whose basis vectors are given 
by the simple roots\footnote{Note that 
despite the fact that normally the lengths of the simple roots are $\sqrt2$, 
here they have length $1$ for simplicity.}
\bea
\alpha_{1} &=&  \left(1,0,0,0,0,0 \right), \nn \\
\alpha_{2} &=&  \left(-\frac{1}{2},\frac{\sqrt{3}}{2},0,0,0,0 \right), \nn \\
\alpha_{3} &=&  \left(0,-\frac{1}{\sqrt{3}},0,-\frac{1}{\sqrt{3}},0,-\frac{1}{\sqrt{3}} \right), \nn \\
\alpha_{4} &=&  \left(0,0,-\frac{1}{2},\frac{\sqrt{3}}{2},0,0 \right), \nn \\
\alpha_{5} &=&  \left(0,0,1,0,0,0 \right), \nn \\
\alpha_{6} &=&  \left(0,0,0,0,-\frac{1}{2},\frac{\sqrt{3}}{2} \right). 
\eea
Then we define the $E_6$ $torus$ $T^6_{E_6} = {\mathbb R}^6/\Lambda_{E_6}$. 
This may seem somewhat a little complicated.
However from the viewpoint of six dimensions, 
it is a highly symmetric space, as we will see. 
We define the root 
\bea
\alpha_{0} &\equiv& 
-\alpha_{1}-2\alpha_{2}-3\alpha_{3}-2\alpha_{4}-\alpha_{5}-2\alpha_{6} \\
&=&\left(0,0,0,0,1,0 \right).\nn 
\eea
This root and the simple roots 
form the $E_6$ extended Dynkin diagram (see figure \ref{e6}). 

The Weyl groups of simple Lie algebras 
and their conjugacy classes are classified in Ref. \cite{Carter:1972}. 
We also take account of the outer automorphisms of $E_6$. 
Due to the symmetry of the Dynkin diagram, 
it has the $\bZ_2$ symmetry 
\beq
g_2 : \ \alpha_1 \lraw \alpha_5, \ \alpha_2 \lraw \alpha_4, \ 
\alpha_i \raw \alpha_i, \ i=3,6,0. 
\eeq  
Note that this action is not included in the Weyl group of $E_6$. 
One can show that the group which is generated from the Weyl group and $g_2$, 
i.e. $\{ \cW, g_2 \}$, is equivalent to $\{ \cW, -\id \}$ \cite{Bourbaki}, 
where $-\id$ acts as  
\beq
-\id : \ \alpha_i \raw -\alpha_i , \ i=0,1, \cdots, 6.
\eeq
There exist the other outer automorphisms in the extended $E_6$ diagram. 
We define one of the elements by 
\beq
g_2^\prime : \ \alpha_5 \lraw \alpha_0, \ \alpha_4 \lraw \alpha_6, \ 
\alpha_i \raw \alpha_i, \ i=1,2,3. 
\eeq  
Because the product of the action $g_2 g_2^\prime$ is 
the $\bZ_3$ element of the automorphism of the extended $E_6$ diagram, 
we see that $g_2$ and $g_2^\prime$ generate all of the outer automorphisms of $E_6$. 
The Weyl group of $E_6$ is generated from 
all the simple roots of the algebra, and it is given by 
\beq
\cW = \{ r_i \ | i=1,\cdots,6 \} = \{ r_i \ | i=2,3,4,5,6,0 \}. 
\eeq
This implies that 
$\{ \cW, g_2 \} = \{ \cW, g_2^\prime \} = \{ \cW, -\id \}$, and 
we conclude that
\beq 
\{ \cW, g_2, g_2^\prime \} = \{ \cW, -\id \}. 
\eeq
From the conjugacy classes of the Weyl groups \cite{Carter:1972} and 
the element $-\id$, 
we can obtain all the $\cN=1$ orbifolds on the $E_6$ torus 
from the automorphism $\{ \cW, -\id \}$ in Table \ref{e6-orbifolds}. 
\begin{table}[t]
\begin{center}
\caption{\it The allowed abelian orbifolds on the $E_6$ lattice, 
which preserve $\cN=1$ supersymmetry. 
Here, $E_6$ is the label of the Coxeter element of the $E_6$ lattice, 
and $E_6(a2)$ and $D_4(a1)$ are the elements of the Carter diagrams 
\cite{Carter:1972}. 
The generators of the $\bZ_N \times \bZ_M$ orbifolds 
are given in the main text and in the appendix.}
\begin{tabular}{|l|l|c|c|c|}
\hline
Orbifold & generators & $\chi$ & $h^{1,1}$ & $h^{2,1}$ \\
\hline \hline
$\bZ_3$ & $(E_6)^4$ or $r_1 r_2 r_4 r_5 r_6 r_0$ & 72 & 36 & 0 \\\hline
$\bZ_4$ & $-\id \times D_4(a1)$ & 48 & 25 & 1 \\\hline
$\bZ_6$-I & $(E_6)^2$ or $E_6(a2)$ & 48 & 25 & 1 \\\hline
$\bZ_6$-II & $r_1 r_2 r_3 r_4 r_5 r_0$ & 48 & 31 & 7 \\\hline
$\bZ_{12}$-I & $E_6 \equiv r_1 r_2 r_3 r_4 r_5 r_6$ & 48 & 25 & 1 \\\hline
$\bZ_2 \times \bZ_2$ & & 24 & 15 & 3 \\\hline
$\bZ_2 \times \bZ_4$ & & 48 & 27 & 3 \\\hline
$\bZ_3 \times \bZ_3$ & & 72 & 36 & 0 \\\hline
\end{tabular}
\end{center}
\label{e6-orbifolds}
\end{table}
The $Z_{12}$-I orbifolds on the $E_6$ lattice have been known in string theory 
\cite{Schellekens:1987ij,Kobayashi:1990fx,Bailin:1999nk}. 
However, we observe that for the other orbifolds on the $E_6$ lattice, 
there are orbifolds with the same Euler numbers and Hodge numbers 
on different lattices \cite{Takahashi:2007qc}, 
except the $\bZ_3 \times \bZ_3$ orbifold. 
This $\bZ_3 \times \bZ_3$ orbifold has interesting properties 
for string model construction, 
and we concentrate on this orbifold in this paper. 

We give the point group elements of the ${\mathbb Z}_3 \times {\mathbb Z}_3$ orbifold 
as follows: 
\bea
\theta &\equiv& r_{1}r_{2}r_{4}r_{5}, \nn \\
\phi &\equiv& r_{5}r_{4}r_{6}r_{0}. 
\eea
They act on the simple roots as 
\begin{align}
\theta:&& \alpha_1 &\raw \alpha_2,&  \alpha_2 &\raw -\alpha_1-\alpha_2,& \ \
\alpha_3 &\raw \alpha_1+\alpha_2+\alpha_3+\alpha_4, \nn \\
&& \alpha_4 &\raw \alpha_5,&  \alpha_5 &\raw -\alpha_4-\alpha_5,&  
\alpha_6 &\raw \alpha_6, \ \ \ \ \ \ \  \alpha_0 \raw \alpha_0, \\ 
\phi:&& \alpha_1 &\raw \alpha_1,&  \alpha_2 &\raw \alpha_2,&  
\alpha_3 &\raw \alpha_3+\alpha_4+\alpha_5+\alpha_6, \nn \\
&& \alpha_4 &\raw -\alpha_4-\alpha_5,& \ \ \alpha_5 &\raw \alpha_4,&  
\alpha_6 &\raw \alpha_0, \ \ \ \ \ \ \  \alpha_0 \raw -\alpha_6-\alpha_0. 
\end{align}
Using the complex coordinates $z_i = x_{2i-1}+ix_{2i}$, with $i=1,2,3$, 
we can rewrite the point group action as
\bea
\theta:& z_i \raw e^{2\pi i v_i} z_i,~~ v = \left(\frac13 ,-\frac13 ,0 \right), \nn \\
\phi:& z_i \raw e^{2\pi i w_i} z_i,~~ w = \left(0,\frac13 ,-\frac13 \right), 
\eea
where $v$ and $w$ are the shift vectors of the point group.
These shifts project out six components of the spinor, 
and leave two chiral spinors, 
$|\pm (\frac12,  \frac12, \frac12, \frac12) \rangle$, invariant.
After the GSO projection, $\cN=1$ supersymmetry is unbroken in four dimensions.

The orbifold action leaves sets of points invariant; i.e., these points differ from 
their orbifold images only by a shift of the torus lattice $\Lambda$.
For the action of $\theta$, $\phi$ or $\theta \phi$, 
these sets appear as two-dimensional fixed tori. 
All the tori fixed by the action of $\theta$ are given by
\beq
\left(0,0,0,0,x,y \right), \ \ \ 
\left( 0,\frac{1}{\sqrt{3}},0,0,x,y \right), \ \ \
\left(0,0,0,\frac{1}{\sqrt{3}},x,y \right),
\eeq
where $x,y \in \bR$. 
We can confirm the number of fixed tori 
using the Lefschetz fixed point theorem \cite{Narain:1986qm}.
The number of fixed tori (\#FT) under the action of $\theta$ is
\beq
\text{\#FT}=\frac{\text{vol}((1-\theta)\Lambda)}{\text{vol}(N)},
\eeq
where $N$ is the lattice normal to the sub-lattice invariant under the action.
Then we have $\text{\#FT}=3$ for the $\theta$-twisted sector. 
Note that by the shift of the $E_6$ root lattice 
the following tori are identified: 
\bea
&(0,0,0,0,x,y) \simeq (0,0,0,0,x,y) - \alpha_3 \simeq
\left(0,\frac{1}{\sqrt{3}},0,\frac{1}{\sqrt{3}},x,y \right).
\eea
This nontrivial structure leads to a smaller number of fixed tori 
than for the ${\mathbb Z}_3 \times {\mathbb Z}_3$ orbifold 
on the factorizable torus $T^2 \times T^2 \times T^2$.

For the action of $\theta \phi^2$, 27 fixed points are invariant. 
Generally, the numbers of fixed $points$ are the same 
for non-factorizable and factorizable orbifolds. 
However, in the orbifold on the $E_6$ torus, 
some of the $\theta \phi^2$ fixed points are not invariant 
under the action of $\theta$ and $\phi$, and 
the states of their linear combinations are left under the projections.
By explicit calculation, we have 15 states, 
which are invariant under $\theta$ and $\phi$, 
and 6 states with a phase of $e^{2\pi i/3}$ and $e^{-2\pi i/3}$ 
under these actions, respectively (see Table \ref{27fix}).

\section{${\mathbb Z}_3 \times {\mathbb Z}_3$ orbifold models on an $E_6$ torus}

\subsection{Gauge embeddings}
Heterotic orbifold models must satisfy some consistency conditions 
required by modular invariance. 
Modular invariance guarantees anomaly cancellation 
in the low energy theory \cite{Vafa:1986wx}.
To satisfy these conditions, we must generally embed the shifts 
in the compact space, $v$ and $w$,
into the gauge shifts, $V$ and $W$ 
\cite{Dixon:1985jw,Dixon:1986jc,Ibanez:1987sn}. 
This condition is described as the level matching condition as follows: 
\bea
& N^\prime [(kV+lW)^2 -(kv+lw)^2] = 0 \bmod2 , \nn \\
& k=0,\cdots,N-1,~~l=0,\cdots,M-1, 
\label{level-match}
\eea
where $N^\prime$ is the order of the twist $\theta^k \phi^l$. 
By inclusion of 
two gauge embeddings,\footnote{Some relations concerning shifts and Dynkin diagrams 
are investigated in Refs. \cite{Hebecker:2003jt} and \cite{Raby:2007yc}.} 
the gauge group $E_8$ can be broken into the GUT gauge groups 
\cite{Georgi:1974sy}, 
but not to $SU(3) \times SU(2) \times U(1)$.
\begin{table}[t]
\begin{center}
\caption{\it The gauge shifts of a $\bZ_3 \times \bZ_3$ orbifold
and the correspondent gauge groups from $E_8$.}
\begin{tabular}{|l|l|}
\hline %
Shifts $V^I$ & Gauge Group  \\
\hline \hline %
$(\frac23,0^7)$, $(\frac13^4,0^4)$, $(\frac13^6,\frac23,0)$ & $SO(14) \times U(1)$  \\[0.3mm]
\hline %
$(\frac13^2,0^6)$, $(\frac13^8)$, $(\frac16^8)$ & $E_7 \times U(1)$  \\[0.3mm]
\hline %
$(\frac13^2,\frac23,0^5)$, $(\frac13^6,0^2)$ & $E_6 \times SU(3)$  \\[0.3mm]
\hline %
$(\frac13^4,\frac23,0^3)$, $(\frac16^7,\frac56)$ & $SU(9)$  \\[0.3mm]
\hline
\end{tabular}
\end{center}
\label{shift-gauge}
\end{table}

The number of $\theta^k \phi^l$-twisted states \cite{Font:1988mk} is given by 
\beq
D(\theta^k \phi^l) = \frac{1}{MN} 
\sum_{t=0}^{N-1} \sum_{s=0}^{M-1} 
\epsilon^{(ks-lt)} 
\tilde{\chi}(\theta^k \phi^l, \theta^t \phi^s) 
\Delta(k,l; t,s),
\label{twist-number}
\eeq
where $\tilde{\chi}$ is the number of the points left simultaneously fixed 
by $\theta^k \phi^l$ and $\theta^t \phi^s$. 
If $\theta^k \phi^l$ leaves some of the coordinates unrotated, 
$\tilde{\chi}$ must be calculated 
using only the sub-lattice that is rotated.
The quantity $\Delta(k,l; t,s)$ is a state-dependent phase, given by 
\bea
\Delta(k,l; t,s) &= P^{(k,l)} \exp\{2\pi i 
[(p+kV+lW)(tV+sW) -(q+ kv+lw)(tv+sw)] \nn \\
&-\frac12 ((kV+lW)(tV+sW)-(kv+lw)(tv+sw)) \}, 
\label{phase-orbi}
\eea
where $P^{(k,l)}$ represents the contribution of the oscillators, 
$p$ is the momentum of the $E_8\times E_8^\prime$ gauge sectors, 
and $q$ is the H-momentum of the twisted states.
Modular invariance for the $\bZ_N \times \bZ_M$ orbifold is satisfied 
by the conditions (\ref{twist-number}) and (\ref{level-match}).

\subsection{Discrete Wilson line}
If we implement Wilson lines as the background in the compact space, 
we can break the gauge group of the models further 
and have a smaller number of generations of matter. 
Discrete Wilson lines are defined 
by the embedding of the shifts in a six dimensional torus 
into the gauge degrees of freedom as  
\beq
\sum_{i=1}^6 l_i \alpha_i \rightarrow \sum_{i=1}^6 l_i a_i^I,~~I=1,\cdots,16, 
\eeq
where $l_i \in \bR$ depends on the location of the fixed points, and $a_i^I \in \Lambda_{E_8\times E_8^\prime}$ is a Wilson line. 
In the untwisted sector, the states that are invariant 
under the action of the Wilson lines survive, 
and this leads to breaking of the gauge group. 
In the twisted sectors, the numbers of degenerate states are reduced, because 
the Wilson lines distinguish the fixed points and tori.

For the ${\mathbb Z}_3 \times {\mathbb Z}_3$ orbifold, 
the shifts of the torus lattice are identified as
$\theta \alpha_{1} = \alpha_{2}$, $\theta \alpha_{4} = \alpha_{5}$ 
and $\phi \alpha_{6} = \alpha_{0}$. 
This means that for the Wilson lines, we have 
\beq
a_{1}^I = a_{2}^I,~~a_{4}^I = a_{5}^I,~~a_{6}^I = a_{0}^I.
\label{wilson1}
\eeq
Moreover, in the $\theta$-twisted sector on the $E_6$ torus, 
the fixed tori that have different shifts are identical: 
\bea
&\left(0,0,0,-\frac{1}{\sqrt{3}},x,y \right) \simeq
\left(0,\frac{1}{\sqrt{3}},0,0,x,y \right).
\eea 
We can read the shifts under the action of $\theta$ from the following: 
\bea
&(1-\theta)\left(0,0,0,-\frac{1}{\sqrt{3}},x,y \right)
= \left(0,0,0,-\frac{1}{\sqrt{3}},x,y \right) + \alpha_4+ \alpha_5, \nn \\
&(1-\theta)\left(0,\frac{1}{\sqrt{3}},0,0,x,y \right)
= \left(0,\frac{1}{\sqrt{3}},0,0,x,y \right) + \alpha_1+ \alpha_2. 
\eea 
It follows that the relation for the Wilson lines is 
\beq
a_4^I +a_5^I =a_1^I+a_2^I, 
\label{wilson2}
\eeq
For a shift $\alpha_3$, we have a Wilson line $a_3$, 
from the definition of $\alpha_0$. 
From (\ref{wilson1}) and (\ref{wilson2}) 
we have only one independent Wilson line, 
$\alpha^I \equiv \alpha^I_i,~ i=0,1,\cdots,6$, for the $E_6$ torus. 
This implies that the orbifold on the $E_6$ torus is highly symmetric space. 

When we include a Wilson line, 
the degeneracy of the fixed tori is reduced from 3 to 1, 
and that of the fixed points of the $\theta \phi^2$-sector is reduced from 27 to 9. 
It seems that a Wilson line breaks the three-family structure of the orbifold, 
and for this reason we do not include Wilson lines in this paper. 
At first sight it seems that the number of states on the 27 fixed points 
is too large for a low energy spectrum. 
However we see below that for some non-standard gauge embeddings, 
the $\theta \phi^2$-twisted sector contains only hidden sector states and singlets. 
Thus we realize three-family structure states from fixed tori.

\subsection{A model with the standard embeddings}

First, we consider an $E_8\times E_8^\prime$ heterotic orbifold model 
obtained from a ${\mathbb Z}_3 \times {\mathbb Z}_3$ orbifold 
on the $E_6$ torus with the standard embeddings. 
The shift vectors that act 
on the gauge sector $E_8\times E_8^\prime$ are given by 
\bea
&v= \left(\frac13, -\frac13, 0\right) ~\raw ~  
V= \left(\frac13,-\frac13,0,0,0,0,0,0\right)\left(0,0,0,0,0,0,0,0\right), \nn \\
&w= \left(0,\frac13, -\frac13\right) ~\raw ~  
W= \left(0,\frac13,-\frac13,0,0,0,0,0\right)\left(0,0,0,0,0,0,0,0\right). 
\eea 
Thus, the level matching condition (\ref{level-match}) is trivially satisfied 
in the standard embedding. 
This corresponds to embedding the spin connection in 
the gauge connection. 

By counting the fixed points and tori,
we easily find the spectrum of this model.
The gauge group is broken to 
\beq
E_6 \times U(1)^2 \times E_8^\prime.
\eeq

There are the twisted sectors 
\beq
\theta,~\theta^2,~\phi,~\phi^2,~\theta\phi,~\theta^2\phi^2,~\theta\phi^2,
\eeq
or $A,~\bar{A},~B,~\bar{B},~C,~\bar{C},~D$, respectively.
These states are distinguished by their H-momenta (listed in Tables 4 and 5). 
There are also three untwisted sector states, $U_1$, $U_2$ and $U_3$. 
The H-momenta of their bosonic states are 
$q_1=(0,1,0,0)$, $q_2=(0,0,1,0)$ and $q_3=(0,0,0,1)$ respectively. 
Then, the matter content of the model is
\begin{center}
\renewcommand{\baselinestretch}{1.5}
\normalsize
\begin{tabular}{rllllllll}
$U_1$, $U_2$, $U_3$: & $3\times \mathbf{27}$, & $A$: & $3\times \mathbf{27}$, 
& $B$: & $3\times \mathbf{27}$,& $C$: & $3\times \mathbf{27}$, \\
$D$: & $15\times \mathbf{27}$, & $\bar{A}$: & $3\times \mathbf{27}$, 
& $\bar{B}$: & $3\times \mathbf{27}$,& $\bar{C}$: & $3\times \mathbf{27}$, \\
\end{tabular}
\end{center}
and singlets. 
Thus, we have 36 generations of matter. 
This number coincides with half of the Euler number $\chi$ of 
this orbifold \cite{Dixon:1985jw,Dixon:1986jc}, 
where $\chi = \sum_{[\theta,\phi]=0} \chi_{\theta,\phi}$. 
Because the generation number of 
the ${\mathbb Z}_3 \times {\mathbb Z}_3$ orbifold model 
on the factorizable torus is 84, 
we find that the generation number is decreased in the non-factorizable model.

\subsection{An $SO(10)$ GUT-like model}

We have seen that 
the ${\mathbb Z}_3 \times {\mathbb Z}_3$ orbifold on the $E_6$ torus
has the phenomenologically interesting feature that it has three fixed tori 
in the $\theta$, $\phi$ and $\theta\phi$-twisted sectors, respectively. 
Here we present an example of models with the $SO(10)$ GUT group. 
Our example is very simple: 
the $E_8\times E_8^\prime$ heterotic string 
from the ${\mathbb Z}_3 \times {\mathbb Z}_3$ orbifold on the $E_6$ torus 
with the gauge embeddings 
\bea
V&=& \left(\frac13,\frac13,\frac23,0,0,0,0,0\right)
\left(\frac13,\frac13,0,0,0,0,0,0\right), \nn \\
W&=& \left(-\frac23,0,0,0,0,0,0,0\right)
\left(\frac13,0,\frac13,\frac13,\frac13,0,0,0\right). 
\eea 
The surviving gauge group in four dimensions is
\beq
SO(10)\times SU(2) \times U(1)^2 \times 
[SU(7) \times U(1)^{2}]^\prime.
\eeq
\begin{table}[bt]
\begin{center}
\caption{\it All of the massless chiral states of 
the $SO(10)\times SU(2) \times U(1)^2 \times 
[SU(7) \times U(1)^{2}]^\prime$ model. 
The entries for the representations are given as 
No.$\times$(Repts.)${}_{Q_1,Q_2,Q_3,Q_4}$.}
\begin{tabular}{|c||c|c||c|c||c|c|}
\hline
 & \multicolumn{2}{c||}{untwisted sector}  
 & \multicolumn{4}{c|}{twisted sector} \\
\cline{2-7}
& & representation & & representation & & representation \\ 
\hline %
visible & $U_3$ & $(\mathbf{16,1})_{3,6,0,0}$ 
& $A$ & $3(\mathbf{16,1})_{{\text -}1,{\text -}2,4,2}$ 
& $\bar{B}$ & $3(\mathbf{10,1})_{{\text -}2,0,{\text -}2,{\text -}8}$\\
sector  & $U_2$ & $(\overline{\mathbf{16}},\mathbf{1})_{{\text -}3,6,0,0}$ 
& $\bar{A}$ & $3(\mathbf{10,1})_{{\text -}2,{\text -}4,{\text -}4,{\text -}2}$ 
& $B$ & $3(\mathbf{1,2})_{2,{\text -}6,2,8}$ \\
        & $U_1$ & $(\mathbf{10,2})_{0,6,0,0}$ 
& $\bar{A}$ & $3(\mathbf{1,2})_{{\text -}2,2,{\text -}4,{\text -}2}$ 
& $\bar{C}$ & $3(\mathbf{1,2})_{2,2,6,{\text -}4}$\\
        & $U_2$ & $(\mathbf{1,2})_{6,6,0,0}$ & & & & \\
        & $U_3$ & $(\mathbf{1,2})_{{\text -}6,6,0,0}$ & & & & \\[0.2mm]
\hline %
messenger& $U_1$ & $\mathbf{1}_{0,{\text -}12,0,0}$ 
& $C$ & $3(\mathbf{1,2})(\mathbf{7})^\prime_{{\text -}2,{\text -}2,0,{\text -}2}$ 
& $D$ & $27\times \mathbf{1}_{0,4,8,18}$ \\
sector   & $U_2$ & $\mathbf{1}_{0,0,12,6}$ & $C$ & $3\times \mathbf{1}_{4,4,6,10}$
& $D$ & $15\times \mathbf{1}_{0,{\text -}8,2,{\text -}6}$ \\
         & & & $A$ & $3\times \mathbf{1}_{{\text -}4,4,4,2}$
& $B$ & $3\times \mathbf{1}_{{\text -}4,0,2,8}$ \\
         & & & $A$ & $3\times \mathbf{1}_{{\text -}4,4,{\text -}8,{\text -}4}$ &  & \\[0.2mm]
\hline %
hidden  & $U_3$ & $(\mathbf{35})^\prime_{0,0,0,{\text -}6}$ 
& $B$ & $3(\mathbf{\bar{7}})^\prime_{{\text -}4,0,2,{\text -}4}$ 
& $D$ & $6(\mathbf{7})^\prime_{0,4,{\text -}4,0}$ \\
sector  & $U_1$ & $(\overline{\mathbf{21}})^\prime_{0,0,6,0}$ 
& $\bar{C}$ & $3(\mathbf{\bar{7}})^\prime_{{\text -}4,{\text -}4,0,2}$ 
& $D$ & $6(\mathbf{\bar{7}})^\prime_{0,4,2,6}$ \\
        & $U_2$ & $(\mathbf{7})^\prime_{0,0,{\text -}6,{\text -}12}$ & & & &  \\
        & $U_3$ & $(\mathbf{7})^\prime_{0,0,0,12}$ & & &  & \\
        & $U_2$ & $(\mathbf{\bar{7}})^\prime_{0,0,{\text -}6,6}$ & & &  & \\[0.2mm]
\hline
\end{tabular}
\end{center}
\label{so(10)-model}
\end{table}
The $U(1)$ directions are 
\begin{eqnarray}
Q_1&=& 6(1,0,0,0,0,0,0,0)(0^8),\nn \\
Q_2&=& 6(0,1,-1,0,0,0,0,0)(0^8),\nn \\
Q_3&=& 6(0^8)(1,1,0,0,0,0,0,0),\nn \\
Q_4&=& 6(0^8)(1,0,1,1,1,0,0,0), 
\end{eqnarray}
and the linear combination 
$Q_A = 3 Q_1-2Q_2 -Q_3 +3Q_4$ is anomalous.  
The matter content is
\beq
3\times (\mathbf{16,1}) + \mbox{vector-like}.
\eeq
Then we have the three-family matter of the $SO(10)$ GUT model.
We also have Higgs states, $(\mathbf{10,1})$, 
but no adjoint Higgs (see Table \ref{so(10)-model}). 

We observe that in the $\theta\phi$-twisted sector, three states 
charged with both the visible and hidden sectors appear.
We call these states the messenger sector, 
because these messenger states have the potential to mediate 
the SUSY-breaking effect through the strong dynamics of the hidden sector.
The running coupling of $SU(7)^\prime$ at the scale $\mu$ is
\beq
\frac{1}{\alpha_{GUT}^\prime} = \frac{1}{\alpha^\prime (\mu)} 
- \frac{b}{2\pi}\ln \left| \frac{M_{GUT}}{\mu} \right|.
\eeq
If all the $(\mathbf{7})$ and $(\mathbf{\bar{7}})$ generate large mass terms,
we have $-b=8$ for $SU(7)^\prime$. 
The confining scale is defined as the scale $\mu$ where $\alpha^\prime(\mu)=1$.
If we have $M_{GUT} = 2 \times 10^{16}$ and $\alpha_{GUT}^\prime = 1/25$, 
the hidden sector scale is estimated as
\beq
\Lambda_{hidden} \sim 1.3 \times 10^8 \mbox{GeV}. 
\eeq
This leads to confinement and gaugino condensation. 
The hidden sector of this model may cause 
gauge-mediated SUSY breaking 
\cite{Affleck:1984xz,Dine:1995ag,Murayama:1995ng,Buchmuller:2006ik,Kim:2007zj}.

\subsection{An $SU(5)$ GUT-like model}
We also construct an $SU(5)$ model. 
We choose the gauge embeddings\footnote{It may seem that 
this shift can be reduced to a more simple form 
by adding lattice vectors. 
However, this generally leads to a different model. 
This freedom with regard to shift vectors is related to 
the discrete torsions \cite{Ploger:2007iq}.} 
\bea
V&=& \left(0,0,0,0,0,\frac13,\frac13,\frac23\right)
\left(0,0,0,0,0,0,\frac13,\frac13\right), \nn \\
W&=& \left(\frac12,\frac12,\frac12,\frac16,\frac56,\frac56,-\frac56,\frac56\right)
\left(\frac23,\frac13,\frac13,0,0,0,0,0\right). 
\eea 
The gauge group of this model is
\beq
SU(5)\times SU(2)_L\times SU(2)_R \times U(1)^2 \times 
[SU(6) \times SU(3) \times U(1)^{}]^\prime.
\eeq
The $U(1)$ directions are 
\begin{eqnarray}
Q_1&=& 3(0,0,0,-1,1,2,0,0)(0^8),\nn \\
Q_2&=& 3(0,0,0,0,0,2,-1,1)(0^8),\nn \\
Q_3&=& 3(0^8)(0,0,0,0,0,0,1,1), 
\end{eqnarray}
and the linear combination 
$Q_A = 8 Q_1 -3Q_2 +2Q_3$ is anomalous.  

We have totally three chiral $\mathbf{10}$s and $\mathbf{\bar{5}}$s of $SO(10)$,
which correspond to the spectrum of the Standard Model. 
Then, the states $(\mathbf{5,1,1})$ and $(\mathbf{\bar{5},1,1})$ 
have just the quantum numbers of the Higgs  
(see Table \ref{su(5)-model}). 
\begin{table}[bt]
\begin{center}
\caption{\it All of the massless chiral states of 
the $SU(5)\times SU(2)_L\times SU(2)_R \times U(1)^2 \times 
[SU(6) \times SU(3) \times U(1)^{}]^\prime$ model.
The entries for the representations are given as 
No.$\times$(Repts.)${}_{Q_1,Q_2,Q_3}$.}
\begin{tabular}{|c||c|c||c|c||c|c|}
\hline
 & \multicolumn{2}{c||}{untwisted sector}  
 & \multicolumn{4}{c|}{twisted sector} \\
\cline{2-7}
& & representation & & representation & & representation \\ 
\hline %
visible & $U_2$ & $(\mathbf{5,1,1})_{0,{\text -}6,0}$ 
& $\bar{A}$ & $3(\mathbf{10,1,1})_{{\text -}2,0,{\text -}2}$ 
& $A$ & $3(\mathbf{1,1,2})_{2,3,{\text -}4}$\\
sector  & $U_3$ & $(\mathbf{\bar{5},1,1})_{6,0,0}$ 
& $A$ & $3(\mathbf{\bar{5},1,1})_{{\text -}4,0,2}$ 
& $A$ & $3(\mathbf{1,1,2})_{2,3,2}$\\
        & $U_2$ & $(\mathbf{10,1,2})_{0,3,0}$ 
& $A$ & $3(\mathbf{1,2,1})_{5,0,2}$ 
& $\bar{A}$ & $3(\mathbf{1,1,2})_{{\text -}2,{\text -}3,{\text -}2}$\\
        & $U_3$ & $(\mathbf{\overline{10},2,1})_{{\text -}3,0,0}$ 
& & & & \\
        & $U_1$ & $(\mathbf{\bar{5},1,2})_{{\text -}6,{\text -}3,0}$ 
& & & & \\
        & $U_1$ & $(\mathbf{5,2,1})_{3,6,0}$ 
& & & & \\
        & $U_1$ & $(\mathbf{1,2,2})_{3,{\text -}3,0}$ 
& & & & \\
        & $U_2$ & $(\mathbf{1,2,1})_{{\text -}9,{\text -}6,0}$ 
& & & & \\
        & $U_3$ & $(\mathbf{1,1,2})_{6,9,0}$ 
& & & & \\[0.2mm]
\hline %
messenger& $U_1$ & $\mathbf{1}_{0,0,{\text -}6}$ 
& $B$ & $3(\mathbf{1,2,1})(\mathbf{1,3})^\prime_{2,4,0}$ 
& $C$ & $3(\mathbf{1,2,1})(\mathbf{6,1})^\prime_{{\text -}3,{\text -}2,{\text -}1}$ \\
sector   & & & $\bar{B}$ & $3(\mathbf{1,1,2})(\mathbf{1,\bar{3}})^\prime_{{\text -}4,{\text -}1,0}$ 
& $A$ & $3\times \mathbf{1}_{{\text -}4,{\text -}6,2}$ \\[0.2mm]
\hline %
hidden  & $U_1$ & $(\mathbf{20,1})^\prime_{0,0,3}$ 
& $D$ & $15(\mathbf{\bar{6},1})^\prime_{{\text -}2,2,{\text -}1}$ 
& $\bar{C}$ & $3(\mathbf{1,\bar{3}})^\prime_{{\text -}6,{\text -}4,{\text -}2}$\\
sector  & $U_3$ & $(\mathbf{15,3})^\prime_{0,0,0}$ 
& $D$ & $6(\mathbf{1,\bar{3}})^\prime_{{\text -}2,2,2}$ & & \\
        & $U_2$ & $(\mathbf{6,\bar{3}})^\prime_{0,0,{\text -}3}$ 
& & & & \\[0.2mm]
\hline
\end{tabular}
\end{center}
\label{su(5)-model}
\end{table}

By counting the hidden sector states charged with $SU(6)^\prime$, 
if pairs of $\mathbf{6}$ and $\mathbf{\bar{6}}$ generate large mass terms, 
we have $-b=10$ for $SU(6)^\prime$. 
The confining scale is estimated as
\beq
\Lambda_{hidden} \sim 5.6 \times 10^9 \mbox{GeV}. 
\eeq
This value is compatible with the hidden sector scale of 
the SUSY-breaking mediation scenario.

\subsection{Three-point functions}
Finally, we consider the Yukawa couplings of 
the ${\mathbb Z}_3 \times {\mathbb Z}_3$ orbifold on the $E_6$ torus. 
To determine the allowed interactions, 
we should take account of H-momentum conservation 
and the space group selection rule. 
In order to determine the general structure of the interactions of the orbifold, 
we ignore its gauge groups in the following. 
The constraint obtained here is common to all models, 
for any gauge embeddings of
the ${\mathbb Z}_3 \times {\mathbb Z}_3$ orbifold on the $E_6$ torus. 

\begin{table}[tb]
\begin{center}
\caption{\it The H-momenta of the bosonic states, the coordinates, and the shifts of 
the $\theta$, $\phi$ and $\theta\phi$-sectors.}
\begin{tabular}{|l|l|l|c||l|l|l|c|}
\hline
 &  & location & shift & &  & location & shift \\
\hline \hline %
$\theta$-sector & $A_{0}$ & $(0,0,0,0,x,y)$ & $0$ & 
$\theta^2$-sector & $\bar{A}_{0}$ & $(0,0,0,0,x,y)$ & $0$ \\ 
$\frac13 (0,1,2,0)$&$A_{1}$ & $(0,0,0,\frac{1}{\sqrt{3}},x,y)$ & $\alpha$ & 
$\frac13 (0,2,1,0)$&$\bar{A}_{1}$ & $(0,\frac{1}{\sqrt{3}},0,0,x,y)$ & $\alpha$ \\ 
&$A_{2}$ & $(0,\frac{1}{\sqrt{3}},0,0,x,y)$ & $-\alpha$ & &
$\bar{A}_{2}$ & $(0,0,0,\frac{1}{\sqrt{3}},x,y)$ & $-\alpha$ \\[0.2mm]
\hline %
$\phi$-sector&$B_{0}$ & $(x,y,0,0,0,0)$ & $0$ & 
$\phi^2$-sector & $\bar{B}_{0}$ & $(x,y,0,0,0,0)$ & $0$ \\ 
$\frac13 (0,0,1,2)$&$B_{1}$ & $(x,y,0,0,0,\frac{1}{\sqrt{3}})$ & $\alpha$ & 
$\frac13 (0,0,2,1)$&$\bar{B}_{1}$ & $(x,y,0,\frac{1}{\sqrt{3}},0,0)$ & $\alpha$ \\ 
&$B_{2}$ & $(x,y,0,\frac{1}{\sqrt{3}},0,0)$ & $-\alpha$ &
&$\bar{B}_{2}$ & $(x,y,0,0,0,\frac{1}{\sqrt{3}})$ & $-\alpha$ \\[0.2mm]
\hline %
$\theta\phi$-sector&$C_{0}$ & $(0,0,x,y,0,0)$ & $0$ &
$\theta^2\phi^2$-sector&$\bar{C}_{0}$ & $(0,0,x,y,0,0)$ & $0$ \\ 
$\frac13 (0,1,0,2)$&$C_{1}$ & $(0,0,x,y,0,\frac{1}{\sqrt{3}})$ & $\alpha$ &
$\frac13 (0,2,0,1)$&$\bar{C}_{1}$ & $(0,\frac{1}{\sqrt{3}},x,y,0,0)$ & $\alpha$ \\ 
&$C_{2}$ & $(0,\frac{1}{\sqrt{3}},x,y,0,0)$ & $-\alpha$ &
&$\bar{C}_{2}$ & $(0,0,x,y,0,\frac{1}{\sqrt{3}})$ & $-\alpha$ \\[0.2mm]
\hline
\end{tabular}
\end{center}
\label{3fix}
\end{table} 
We identify the individual states in each twisted sector, 
$A,~\bar{A},~B,~\bar{B},~C,~\bar{C}$ and $D$, 
by its space group. 
For example, in the $\theta$-twisted sector (labeled $A$),
we can assign a state which is localized at $f_\theta$ 
a space group element $(\theta,l)$. 
Here, $l$ is the shift of the state, defined by 
\beq
l=(\id-\theta)f_{\theta}  \pmod{(\id-\theta)\Lambda},
\eeq
where we have $l \in \Lambda$, by definition. 
The space group corresponds to the boundary condition of the closed string as
$X(2\pi)= (\theta,l)X(0)= \theta X(0) + l$.
Therefore, the boundary conditions for the string interaction of three twisted state is 
\begin{eqnarray}
X(2\pi)&=& (\theta_1,l_1)(\theta_2,l_2)(\theta_3,l_3)X(0),\nn\\
&=& \theta_1 \theta_2 \theta_3 X(0) 
+ l_1 + \theta_1 l_2 + \theta_1 \theta_2 l_3.
\end{eqnarray}
Then, we have the space group selection rules 
\bea
&&\theta_1 \theta_2 \theta_3 = I, \nn\\
&&l_1 + \theta_1 l_2 + \theta_1 \theta_2 l_3 = 0.
\eea
We call the former equation the point group selection rule. 
Because the shifts are defined up to the sublattice $(\id-\theta)\Lambda$, 
the latter condition simplifies to
\beq
l_1 + l_2 + l_3 = 0  \pmod{\sum_{i=1}^3 (\id-\theta_i)\Lambda}.
\label{space-select}
\eeq
Actually, these are the conditions required for non-zero correlation functions.
The three-point interactions consistent with the point group selection rule 
and the H-momentum conservation 
\cite{Bailin:1987xm, Kobayashi:1990fx, Bailin:1999nk} are
\bea
&&U_1 U_2 U_3,~~
U_3 A \bar{A},~~
U_1 B \bar{B},~~
U_2 C \bar{C},~~
D D D, \nn \\
&&
\bar{A} B D,~~
A C D,~~
\bar{B} \bar{C} D,~~
A B \bar{C},~~
\bar{A} \bar{B} C, 
\eea
where $U_1$, $U_2$ and $U_3$ are untwisted sector states.   
We should also take account of the constraint from (\ref{space-select}).

\begin{table}[tb]
\begin{center}
\caption{\it The coordinates and the shifts 
of the 27 states in the $\theta\phi^2$-sector. 
The H-momentum is $\frac13 (0,1,1,2)$ for all states. 
In this table, we use abbreviations in the coordinates as follows: 
$+ = (\frac{1}{\sqrt3},0)$, $- = (-\frac{1}{\sqrt3},0)$, 
$a = (\frac13,0)$, $b = (-\frac16,\frac{1}{2\sqrt3})$ 
and $c = (-\frac16,-\frac{1}{2\sqrt3})$. 
Also, the symbols indicate that the quantity in question is $-1$ times 
that without the bar, i.e., $\bar{a}\equiv -a$. 
For example we have the coordinate 
$(a,b,c)= (\frac13,0,-\frac16,\frac{1}{2\sqrt3},-\frac16,-\frac{1}{2\sqrt3})$.
Note that under the identification of $\theta\phi^2$, 
we observe $(a,a,a)=(b,b,b)=(c,c,c)$, etc.} 
\begin{tabular}{|@{\ }l@{\ }|@{\ }l@{\ }|l||@{\ }l@{\ }|@{\ }l@{\ }|l||@{\ }l@{\ }|@{\ }l@{\ }|l|}
\hline
 & location & shift & 
 & location & shift & 
 & location & shift  \\
\hline \hline %
$D_{00}$ & $(0,0,0)$ & $0$ & 
$D_{01}$ & $(-,+,0)$ & $\alpha_2-\alpha_4$ & 
$D_{02}$ & $(+,-,0)$ & $-\alpha_2-\alpha_4$ \\
$D_{10}$ & $(-,0,0)$ & $\alpha_2$ & 
$D_{11}$ & $(0,-,0)$ & $\alpha_4$ & 
$D_{12}$ & $(0,0,-)$ & $-\alpha_2-\alpha_4$ \\
$D_{20}$ & $(+,0,0)$ & $-\alpha_2$ & 
$D_{21}$ & $(0,+,0)$ & $-\alpha_4$ & 
$D_{22}$ & $(0,0,+)$ & $\alpha_2${\scriptsize+}$\alpha_4$ \\[0.2mm]
\hline %
$D_{00}^{'}$ & $(a,a,a)$ & $-\alpha_3$ & 
$D_{01}^{'}$ & $(a,b,c)$ & $-\alpha_3+\alpha_2-\alpha_4$ & 
$D_{02}^{'}$ & $(a,c,b)$ & $-\alpha_3-\alpha_2+\alpha_4$ \\
$D_{10}^{'}$ & $(b,a,a)$ & $-\alpha_3+\alpha_2$ & 
$D_{11}^{'}$ & $(a,b,a)$ & $-\alpha_3+\alpha_4$ & 
$D_{12}^{'}$ & $(a,a,b)$ & $-\alpha_3-\alpha_2-\alpha_4$ \\
$D_{20}^{'}$ & $(c,a,a)$ & $-\alpha_3-\alpha_2$ & 
$D_{21}^{'}$ & $(a,c,a)$ & $-\alpha_3-\alpha_4$ & 
$D_{22}^{'}$ & $(a,a,c)$ & $-\alpha_3+\alpha_2+\alpha_4$ \\[0.2mm]
\hline %
$D_{00}^{''}$ & $(\bar{a},\bar{a},\bar{a})$ & $\alpha_3$ & 
$D_{01}^{''}$ & $(\bar{a},\bar{b},\bar{c})$ & $\alpha_3+\alpha_2-\alpha_4$ & 
$D_{02}^{''}$ & $(\bar{a},\bar{c},\bar{b})$ & $\alpha_3-\alpha_2+\alpha_4$ \\ 
$D_{10}^{''}$ & $(\bar{b},\bar{a},\bar{a})$ & $\alpha_3+\alpha_2$ & 
$D_{11}^{''}$ & $(\bar{a},\bar{b},\bar{a})$ & $\alpha_3+\alpha_4$ & 
$D_{12}^{''}$ & $(\bar{a},\bar{a},\bar{b})$ & $\alpha_3-\alpha_2-\alpha_4$ \\ 
$D_{20}^{''}$ & $(\bar{c},\bar{a},\bar{a})$ & $\alpha_3-\alpha_2$ & 
$D_{21}^{''}$ & $(\bar{a},\bar{c},\bar{a})$ & $\alpha_3-\alpha_4$ & 
$D_{22}^{''}$ & $(\bar{a},\bar{a},\bar{c})$ & $\alpha_3+\alpha_2+\alpha_4$ \\[0.2mm] 
\hline
\end{tabular}
\end{center}
\label{27fix}
\end{table}

First, we consider the coupling of $\bar{A}BD$. 
The sum of the sublattices generated by the action of $\theta$, $\phi$ and $\theta\phi^2$ is
\bea
&&(\id-\theta)\Lambda + (\id-\phi)\Lambda + (\id-\theta\phi^2)\Lambda \nn \\ 
&&=\{ 
\alpha_1-\alpha_2,~ \alpha_2-\alpha_4,~ \alpha_4-\alpha_5,~
\alpha_4-\alpha_6,~ \alpha_6-\alpha_0,~ 3\alpha_2 \}, 
\label{sublattice1}
\eea
Therefore, in the space group selection rule of $\bar{A}BD$, we can identify 
\beq
\alpha \equiv \alpha_1 =\alpha_2 =\alpha_4 =\alpha_5 =\alpha_6 =\alpha_0,~~
3\alpha = 3\alpha_3 \simeq 0,
\label{3-point-root}
\eeq
up to sublattices.
The last identity is obtained from the definition of $\alpha_0$. 
However, we have $\alpha \not= \alpha_3$. 
The space group elements of the three states 
in the $\theta$-twisted sector are given by 
$(\theta,0)$, $(\theta,\alpha)$, $(\theta,-\alpha)$, respectively, 
and this is similar for the $\phi$ and $\theta\phi$-sectors. 
The shifts and the coordinates of these states are listed in Table \ref{3fix},
and the states in the $\theta\phi^2$-sectors
are listed in Table \ref{27fix}. 
According to the space group selection rule, 
$\bar{A}BD$ couplings are allowed for 
$D$ states with no $\alpha_3$ shift, that is $D_{ij}$, $i,j=0,1,2$. 
Couplings with $D_{ij}^{'}$ and $D_{ij}^{''}$ are forbidden. 
Then, in the interactions of $AB\bar{C}$, the coupling 
\beq
\bar{A}_i B_j D_{kl},~~i+j+k=0 \pmod3,~l=0,1,2,
\label{degenerate1}
\eeq 
can take non-zero values.
The interaction is allowed for any $l$, 
and such interactions do not appear in factorizable orbifold models. 
In this sence, $\bar{A}BD$ couplings include 
nontrivial structures\footnote{We observed similar phenomena 
in ${\mathbb Z}_2 \times {\mathbb Z}_2$ orbifolds 
on non-factorizable lattices \cite{Forste:2006wq}.}. 
The same is true for the couplings of $A C D$ and $\bar{B} \bar{C} D$. 
This interaction is generated because 
three fixed points of $D$ are on the same fixed tori of $\bar{A}$ or $B$.
Some of these are not contact interactions, and 
are generated by the effect of the world sheet instanton \cite{Bailin:1999nk}. 
Then, the couplings of the three-point interactions 
are suppressed by the instanton effect $\varepsilon$. 
The suppression factor is  related to the distances between 
the three associated fixed points, and 
the dominant contribution is from the instanton effects 
on the minimum of this distance. 
Taking this into consideration, we can estimate the 
coupling of the three-point functions. 
Suppose that $B_j$ and $D_{2l}$ are interpreted as the three-family matter states 
and $\bar{A}_i$ is the Higgs $H_i$ $(i=0,1,2)$. 
Then, the following interactions are allowed, 
\beq
\bar{A}_0 B_1 D_{2l},~~\bar{A}_1 B_0 D_{2l},~~\bar{A}_2 B_2 D_{2l},~~l=0,1,2. 
\eeq
Then the Yukawa matrix is 
\begin{equation}
Y = \left(
\begin{array}{ccc}
H_1 & \varepsilon H_1 & \varepsilon H_1 \\
\varepsilon H_0 & \varepsilon H_0 & H_0 \\
\varepsilon H_2 & H_2 & \varepsilon H_2 
\end{array}
\right),
\label{yukawa}
\end{equation}
with an order 1 factor from quantum corrections. 
Thus, in this case, we can realize the Yukawa matrix with mixing. 
It is also notable that three-generation structure naturally arises 
in the $Z_3 \times Z_3$ orbifold on the $E_6$ torus.

Next, we consider the couplings of $AB\bar{C}$. 
The sum of the sublattices
\bea
&&(\id-\theta)\Lambda + (\id-\phi)\Lambda + (\id-\theta\phi)\Lambda \nn \\ 
&&=\{ 
\alpha_1-\alpha_2,~ \alpha_2-\alpha_4,~ \alpha_4-\alpha_5,~
\alpha_4-\alpha_6,~ \alpha_6-\alpha_0,~ 3\alpha_2
\}
\eea
is the same as in (\ref{sublattice1}), and 
therefore we can use the relations (\ref{3-point-root}) in this case.
Then, the coupling 
\beq
A_i B_j \bar{C}_k,~~i+j+k=0 \pmod3,
\eeq 
can take non-zero values.
Assuming that $A_i$ and $B_j$ are the three-family matter states 
and $\bar{C}_k$ is the Higgs $H_k$ $(k=1,2,3)$, 
the Yukawa matrix is given by
\begin{equation}
Y = \left(
\begin{array}{ccc}
H_0 & H_2 & H_1 \\
H_2 & H_1 & H_0 \\
H_1 & H_0 & H_2
\end{array}
\right),
\label{yukawa1}
\end{equation}
with an order 1 factor from quantum corrections. 
These are contact interactions and 
are not suppressed by $\varepsilon$. 
We have the same structures for $\bar{A} \bar{B} C$.

Finally, we consider the interactions of $DDD$ type. 
One complication in this case is that 
the fixed points in Table \ref{27fix} are not invariant 
under the actions of $\theta$ and $\phi$. 
For this reason we should take linear combinations of these states 
in order to obtain eigenstates of the orbifold actions given by 
\bea
D_{j}^{'k} &\equiv& D_{j0}^{'} + \omega^k D_{j1}^{'} +\omega^{2k} D_{j2}^{'}, \nn \\
D_{j}^{''k} &\equiv& D_{j0}^{''} + \omega^k D_{j1}^{''} +\omega^{2k} D_{j2}^{''},  
\eea
where $\omega= e^{2\pi/3i}$ and j,k=0,1,2. 
The interactions allowed by the selection rule (\ref{space-select}) 
in terms of $\alpha_3$ are of 
$DDD$, $D^{'}D^{'}D^{'}$, $D^{''}D^{''}D^{''}$ and $D^{}D^{'}D^{''}$ types. 
It is straightforward to show that the following self-interactions can exist: 
\beq
D_{ij}D_{ij}D_{ij},~D_{ij}^{'}D_{ij}^{'}D_{ij}^{'},~
D_{ij}^{''}D_{ij}^{''}D_{ij}^{''}, 
\eeq
where $i,j=0,1,2$, and we do not sum over their indices. 
These are contact interactions, and hence the couplings are order 1. 
There are other couplings for 
\beq
D_{0i}D_{1j}D_{2k},~~D_{0}^{'i}D_{1}^{'j}D_{2}^{'k},~~
D_{0}^{''i}D_{1}^{''j}D_{2}^{''k},~~i+j+k=0 \pmod3. \\
\label{D012}
\eeq
However these are not contact interactions,  
and the minimum distance associated with the instanton effect 
is the same as for the interactions above.  
If the state $D_{0i}$ for $i=0,1,2$ is Higgs, 
we have a Yukawa matrix (\ref{yukawa1}) with the suppression factor $\varepsilon$. 
We also have non-zero couplings for
\beq
D_{il}D_{j}^{'1}D_{k}^{''2},~~D_{il}D_{j}^{'2}D_{k}^{''1},~~
i+j+k=0 \pmod3,~l=0,1,2,\\
\label{degenerate2}
\eeq 
Here, we again observe degenerate interactions for $l$. 
These couplings are generated by instantons, 
and are suppressed by factor of $\varepsilon$. 
In this case, the minimum distances for the instanton effect are the same for all.

We have considered the allowed couplings from the space group selection rules here. 
We have obtained the interactions (\ref{degenerate1}) and (\ref{degenerate2}),
and these are new structures in the non-factorizable orbifold models.
The coupling of $\bar{A}BD$ can generate mixing terms between flavors. 
The 27 fixed states in the $\theta\phi^2$-sector 
are divided into interactions among the three flavors. 
For further model construction, 
there is a possibility to realize three-family states from this sector 
with flavor mixing interactions.

\section{Conclusion}

Taking consideration of the Weyl group and outer automorphisms, 
we classified the orbifolds on the $E_6$ lattice, 
which preserve $\cN=1$ supersymmetry, 
and found ${\mathbb Z}_4$, ${\mathbb Z}_2 \times {\mathbb Z}_2$,  
${\mathbb Z}_2 \times {\mathbb Z}_4$ and ${\mathbb Z}_3 \times {\mathbb Z}_3$ 
orbifolds are allowed. 
The Euler numbers and Hodge numbers of these orbifolds are listed 
in Table \ref{e6-orbifolds}.

In particular we considered the heterotic string models on 
the ${\mathbb Z}_3 \times {\mathbb Z}_3$ orbifold on the $E_6$ torus in detail, 
because the ${\mathbb Z}_3 \times {\mathbb Z}_3$ orbifold includes three fixed tori 
in the $\theta$, $\phi$ and $\theta\phi$-twisted sectors, respectively, 
and easily leads to three-family spectra. 
Such a six-dimensional orbifold with three degenerate fixed tori (or points), 
which leads to $\cN=1$ models, was not previously known. 
We gave examples of $\cN=1$ three-family models 
from the ${\mathbb Z}_3 \times {\mathbb Z}_3$ orbifold on the $E_6$ torus. 
Our assumption is quite simple, i.e. 
compactification on the orbifold with two gauge embeddings in the models. 
We found that these constructions lead to simple spectra, 
owing to the small number of fixed tori. 
The models have strongly coupled sectors in the low energy 
and messenger-like states charged 
with both the hidden and visible sector gauge groups.
In order to have these strongly coupled sectors, 
the number of flavors should be sufficiently small. 
Whereas there are often too many massless states in heterotic orbifold models, 
the numbers of twisted states on non-factorizable orbifolds can be 
smaller by factors of two or three than those of 
the corresponding factorizable orbifold. 
However we have less freedom regarding the moduli space. 
These facts imply that this $E_6$ orbifold is quite symmetric 
from the viewpoint of the six-dimensional space, 
and such a symmetric space is natural for a compact space. 
The phenomenological problem of the GUT-like models is that 
they do not include adjoint Higgs that cause GUT group breaking.
This is a notorious obstacle for the level $k=1$ construction of heterotic strings. 
In this paper, we consider models with no Wilson lines, 
because it seems that Wilson lines break the structure of 
the degenerate three fixed tori. 
However, the inclusion of Wilson lines may lead to other three-family models 
whose families are generated from different twisted sectors. 
If we introduce continuous Wilson lines 
\cite{Ibanez:1987xa,Font:1988mm,Forste:2005rs,Forste:2005gc},
we may be able to realize models with rank-reduced gauge groups. 

We also considered the general properties of the three-point interactions 
allowed by the space group selection rules. 
To this time, it has only been showed that 
non-prime $\bZ_N$ orbifolds, where $N$ is not a prime number, 
can generate flavor mixing terms at the tree level. 
We found that the selection rules for three-point interactions 
of the world sheet instanton effect differ from those for the factorizable model, 
and this leads to non-trivial structures of the mass matrices. 
For example, 27 states in the $\theta\phi^2$-sector  
are divided into 3 flavors with respect to their interactions, 
and we observe three-family interactions with mixing. 
Such a structure of the interactions can generate texture, 
in other words, hidden symmetries of the mass matrices. \\

\subsection*{Acknowledgement}

I would like to thank T. Kobayashi, T. Kimura and M. Ohta for valuable discussions. 
I also thank F. Ploeger, S. Ramos-Sanchez and P. Vaudrevange 
for pointing out errors in the first version of this paper. 
K.~T.\/ is supported by the Japan Society for the Promotion of Science.

\appendix

\subsection*{Appendix}

\subsection*{$\bZ_N \times \bZ_M$ Orbifolds on the $E_6$ Lattice}
Here, we give explicit representations 
for the $\bZ_N \times \bZ_M$ orbifolds on the $E_6$ lattice 
listed in Table \ref{e6-orbifolds}. 

\begin{itemize}

\item {$\bZ_2 \times \bZ_2$ orbifold} \\
For the $\bZ_2 \times \bZ_2$ orbifold, 
it would be convenient to use the following basis for the $E_6$ lattice: 
\begin{align}
\alpha_{1} & =  \left(1,-1,0,0,0,0 \right),\nn \\
\alpha_{2} & =  \left(0,1,-1,0,0,0 \right),\nn  \\
\alpha_{3} &=  \left(0,0,1,-1,0,0 \right),  \nn \\
\alpha_{4} &=  \left(0,0,0,1,-1,0 \right), \nn \\
\alpha_{5} &=  \left(0,0,0,1,1,0 \right), \nn \\
\alpha_{6} &=  \left(-\frac12,-\frac12,-\frac12,\frac12,\frac12,-\frac{\sqrt{3}}{2} \right). 
\label{so-base}  
\end{align}
The point group elements are given by 
\bea
\theta &=& \text{diag}(-1,-1,-1,-1,1,1),\nn \\
\phi &=& \text{diag}(1,1,-1,-1,-1,-1),
\eea
and the Euler number and Hodge numbers are given in Table \ref{e6-orbifolds}.
We can also employ a different representation in the same conjugacy class, 
given by 
\bea
\theta &=& -\id r_1 r_3, \nn \\
\phi &=& -\id r_6 r_0. 
\eea
This leads to the same result.

\item $\bZ_2 \times \bZ_4$ orbifold \\
The point group elements of the $\bZ_2 \times \bZ_4$ orbifold are given by 
\bea
\theta &=& \left(
\begin{array}{cccccc}
0 &-1 & 0 & 0 & 0 & 0 \\
1 & 0 & 0 & 0 & 0 & 0 \\
0 & 0 & 0 & 1 & 0 & 0 \\
0 & 0 &-1 & 0 & 0 & 0 \\
0 & 0 & 0 & 0 & 1 & 0 \\
0 & 0 & 0 & 0 & 0 & 1 
\end{array}
\right), \nn \\
\phi &=& \text{diag}(1,1,-1,-1,-1,-1), 
\eea
in the basis (\ref{so-base}),
and the Euler number and Hodge numbers are given in Table \ref{e6-orbifolds}.

\item $\bZ_3 \times \bZ_3$ orbifold \\
This orbifold is explained in detail in the main text. 

\end{itemize}

\end{document}